\documentclass[twocolumn,superscriptaddress]{revtex4-1}
\usepackage[utf8]{inputenc}
\usepackage[version=3]{mhchem}
\usepackage{enumitem}
\usepackage{multirow}
\usepackage{xcolor}
\usepackage{graphicx}
\usepackage{amssymb}
\setcitestyle{super}

\usepackage{mathtools}

\newcommand{\figref}[1]{Figure~\ref{#1}}
\newcommand{\tabref}[1]{Table~\ref{#1}}

\definecolor{light-gray}{gray}{0.95}

\DeclareRobustCommand*{\ocite}[1]{%
	\begingroup
	\romannumeral-`\x 
	\setcitestyle{numbers,open={},close={}}%
	\cite{#1}%
	\endgroup   
}

\begin{document}
	
\title{How will quantum computers provide an industrially relevant computational advantage in quantum chemistry?}

\author{Vincent E. Elfving}
\email{vincent.elfving@quandco.com}
\affiliation{Qu \& Co B.V., Amsterdam, The Netherlands}

\author{Benno W. Broer}
\affiliation{Qu \& Co B.V., Amsterdam, The Netherlands}

\author{Mark Webber}
\affiliation{IQT, University of Sussex, United Kingdom}

\author{Jacob Gavartin}
\affiliation{Schrödinger Inc., 20 Station Road, Cambridge, Cambridgeshire
	CB1 2JD, United Kingdom}

\author{Mathew D. Halls}
\affiliation{Schrödinger Inc., 10201 Wateridge Circle, Suite 220
	San Diego, CA 92121, United States}

\author{K. Patrick Lorton}
\affiliation{Schrödinger Inc., 120 West 45th St, 17th Fl., New York, NY 10036, United States}

\author{Art D. Bochevarov}
\email{art.bochevarov@schrodinger.com}
\affiliation{Schrödinger Inc., 120 West 45th St, 17th Fl., New York, NY 10036, United States}

\date{September 18, 2020}

\begin{abstract}
	Numerous reports claim that quantum advantage, which should emerge as a direct consequence of the advent of quantum computers, will herald a new era of chemical research because it will enable scientists to perform the kinds of quantum chemical simulations that have not been possible before. Such simulations on quantum computers, promising a significantly greater accuracy and speed, are projected to exert a great impact on the way we can probe reality, predict the outcomes of chemical experiments, and even drive design of drugs, catalysts, and materials. In this work we review the current status of quantum hardware and algorithm theory and examine whether such popular claims about quantum advantage are really going to be transformative. We go over subtle complications of quantum chemical research that tend to be overlooked in discussions involving quantum computers. We estimate quantum computer resources that will be required for performing calculations on quantum computers with chemical accuracy for several types of molecules. In particular, we directly compare the resources and timings associated with classical and quantum computers for the molecules H$_2$ for increasing basis set sizes, and Cr$_2$ for a variety of complete active spaces (CAS) within the scope of the CASCI and CASSCF methods. The results obtained for the chromium dimer enable us to estimate the size of the active space at which computations of non-dynamic correlation on a quantum computer should take less time than analogous computations on a classical computer. The transition point should occur at around $19 \leq N \leq 34$, for CAS of the type $(N, N)$, under the assumption of the much-researched surface code. This is significantly smaller than the active spaces discussed in the context of quantum advantage in prior publications. Using this result, we speculate on the types of chemical applications for which the use of quantum computers would be both beneficial and relevant to industrial applications in the short term. 
\end{abstract}

\maketitle

\section{Introduction}

It has often been predicted that quantum chemistry will greatly benefit from the use of future quantum computers, and therefore multiple quantum computational algorithms have been discussed in the context of chemical applications.\cite{Aspuru-Guzik2005, Lanyon-NC-2010, Kandala2017,  Reiher2017ElucidatingComputers, PhysRevLett.120.110501, Cao2019QuantumComputing} A driving force behind attempting to do electronic structure theory on quantum computers is a reduction of the exponential scaling of some of the theory's methods to a polynomial one.\cite{Aspuru-Guzik2005, Kassal18681, Cao2019QuantumComputing} Of special interest are applications which would not only enjoy the significant speedup provided by quantum computers, but would be propelled from the category of computationally ``impossible'' to that of feasible.\cite{Reiher2017ElucidatingComputers} 

Multiple surveys discuss how to implement electron structure theory on quantum computers, \cite{Whitfield2011d,McArdle2018h,Cao2019QuantumComputing} while other surveys have provided some overview of chemistry-related applications, which are expected to benefit from quantum advantage.\cite{Bauer2020QuantumScience,Cao2018i} Some papers have focused on the quantum resources required for solving a specific chemical problem \cite{wecker2014, Reiher2017ElucidatingComputers, Kivlichan2019, burg2020quantum} and others have proposed novel quantum algorithms to improve those resource requirements. \cite{PhysRevLett.120.110501,Svore2014,Kassal18681,Dallaire-Demers2018b} While the mentioned reviews are very informative, we nevertheless feel that there still exists a disconnect between the quantum chemistry and quantum computing communities in the way they use their terminology, set goals for demonstrating quantum advantage, or choose potential practical applications. The present work is intended to bridge this gap.

\begin{figure}
	\includegraphics[width=0.99\linewidth]{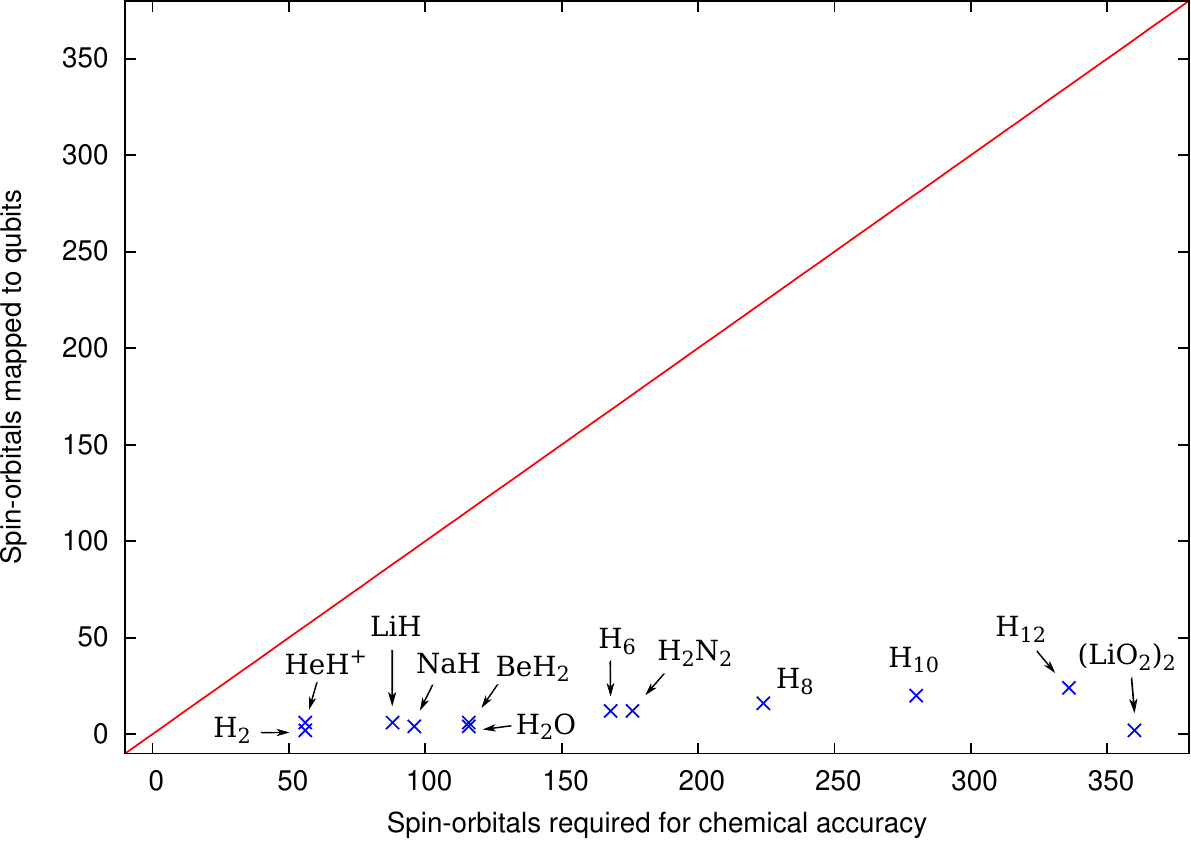}
	\caption{A comparison between the number of spin-orbitals required for achieving chemical accuracy on the ground state and the number of spin-orbitals mapped to qubits in a few actual quantum computing calculations conducted to date. Here it is assumed that the cc-pVTZ basis set will be sufficient to achieve chemical accuracy, if used as the largest basis set in an extrapolation scheme. For the details on the quantum computing data see Table~\ref{nisq_experiments}. The conversion 1 qubit = 1 spin-orbital\cite{dirac1927, Jordan1928} was used for all molecular systems except for $\text{H}_{6}$, $\text{H}_{8}$, $\text{H}_{10}$, $\text{H}_{12}$, and $\text{H}_{2}\text{N}_{2}$ for which the conversion 1 qubit = 2 spin-orbitals was applied in accordance with Ref.~\ocite{arute2020hartreefock}.  The red line $y=x$ shows what the number of spin-orbitals mapped to qubits needs to be in order to satisfy the demand for chemical accuracy.
		\label{fig:basis}}
\end{figure}

In order to put quantum computing algorithms on a well-charted map of classical computational chemistry algorithms we first quantify the limits of conventional state-of-the-art molecular chemistry simulations. We provide insight into the requirements for meaningful quantum advantage, and investigate the quantum resources and estimated runtime needed to realize such quantum advantage. As part of our investigation, we revisit an often cited reference problem for early quantum advantage, a nitrogen fixation catalyzed by FeMo-co (short for FeMo-cofactor),\cite{Reiher2017ElucidatingComputers} and argue that a breakthrough in FeMo-co research promises to be more complicated than perhaps anticipated. We investigate the particular strengths and weaknesses of quantum computational chemistry to provide direction for research towards early relevant quantum advantage and propose criteria for, and dimensions of, molecular systems on which this advantage can be exercised.

\section{Accuracy vs. precision}
The most useful measure of accuracy for applications involving chemical reactions is chemical accuracy, i.e. an error of less than 1 kcal/mol with respect to the hypothetical ``exact'' energy or an experimental measurement fully devoid of error. Chemical accuracy is a desirable target because calculations capable of achieving it would rival the accuracy of measurements attainable in a chemical laboratory.

It is important, when talking about calculations on a quantum computer, to distinguish accuracy (computational error with respect to an experimental measurement) from precision (computational error with respect to a computational reference, for example, a sufficiently accurate result obtained with a large basis set). The mixing up of these terms is still widespread in the quantum computational community. 

Thus, over the last few years several papers,\cite{Peruzzo2014, Kandala2017, OMalley2016, PhysRevX.8.011021, Ryabinkin-JCTC-2018, Armaos2019b, Elfving2020SimulatingDevice, arute2020hartreefock} (including one by one of the authors of the present work) described quantum computing experiments that, according to the authors, reached chemical accuracy. In reality, they reached chemical precision: an error of at most 1 kcal/mol compared to the exact solution typically provided by the combination of the full configuration interaction (FCI) method and a very small basis set that was used as a reference. It is important that in the future the quantum computational community use these terms correctly, because an answer computed with a 1 kcal/mol precision may be useless for explaining or predicting chemical reactivity if the level of theory used as a reference does not allow a similar level of accuracy.  

Of course, matching an FCI energy value is not necessarily a practically useful achievement. It means attaining a good precision but not necessarily a high accuracy. In this work we will argue that while developing quantum computing algorithms it would be more useful to target chemical accuracy obtained with an appropriate combination of the method (not necessarily FCI) and the basis set that are capable of yielding such accuracy.

Let us consider the relationship between the energy and the basis set size, as it is intimately linked with the problem of achieving chemical accuracy. It is well known that configuration interaction (including FCI) energies obtained with Gaussian basis sets converge to the ``exact'' energy very slowly.\cite{Valeev-ARCC-2006, Helgaker-MP-2008} Therefore one needs Gaussian basis sets of at least quadruple-$\zeta$ quality (where $\zeta$ is the number of contracted functions per atomic orbital) to achieve chemical accuracy with pure, non-extrapolated FCI calculations.\cite{Valeev-ARCC-2006} For the H$_2$ and He molecules, for example, one does not achieve chemical accuracy at the FCI level with the cc-pVTZ basis set (28 basis functions or 56 spin-orbitals).\cite{Lee-Park-JCP-2000} One has to use either cc-pVQZ or both cc-pVDZ and cc-pVTZ in an extrapolation scheme for H$_2$, whereas for He at least the cc-pVTZ (14 functions or 28 spin-orbitals) and cc-pVQZ (30 functions or 60 spin-orbitals) basis set energies are required as inputs to an extrapolation scheme.\cite{Lee-Park-JCP-2000} 

Figure~\ref{fig:basis} presents a comparison between the number of spin-orbitals required for achieving chemical accuracy for the ground state and the equivalent number of spin-orbitals used to date in representative, actual quantum computer calculations (for a more complete summary see Table~\ref{nisq_experiments}). The diagram makes it clear that the number of qubits that have been used in quantum chemical computations to date is significantly smaller than that necessary to achieve chemical accuracy, for all molecules studied. 

\begin{table*}[ht]
	\centering
	\begin{tabular}{|c|c|c|c|c|c|}
		\hline
		\textbf{Reference} & \textbf{Year} & \textbf{Max \# qubits} & \textbf{Systems} & \textbf{Platform} & \textbf{Methods} \\ \hline
		Peruzzo et al. \cite{Peruzzo2014} & 2013 & 2 & $\text{HeH}^{+}$ & Silicon Photonic & VQE-UCC \\ \hline
		Shen et al. \cite{1506.00443} & 2015 & 2 & $\text{HeH}^{+}$ & Trapped ion & VQE-UCC \\ \hline
		Google \cite{OMalley2016} & 2015 & 2 & $\text{H}_2$ & Superconducting & VQE-UCC \\ \hline
		Santagati et al. \cite{1611.03511} & 2016 & 2 & $\text{H}_2$, $\text{H}_3$, $\text{H}_{3}^{+}$, $\text{H}_4$ & Silicon photonic & IPEA, VQE-UCC \\ \hline
		IBM \cite{Kandala2017} & 2017 & 6 & \begin{tabular}[c]{@{}c@{}}$\text{H}_2$, LiH, $\text{BeH}_2$,\\ Heisenberg model\end{tabular} & Superconducting & \begin{tabular}[c]{@{}c@{}}Hardware-efficient\\ VQE\end{tabular} \\ \hline
		Berkeley \cite{PhysRevX.8.011021} & 2017 & 2 & $\text{H}_2$ (excited states) & Superconducting & \begin{tabular}[c]{@{}c@{}}Hardware-specific\\ VQE\end{tabular} \\ \hline
		Hempel et al. \cite{PhysRevX.8.031022} & 2018 & 3 & $\text{H}_2$, LiH & Trapped-ion & VQE-UCC \\ \hline
		IBM \cite{Kandala2019} & 2018 & 4 & \begin{tabular}[c]{@{}c@{}}Quantum magnetism\\ $\text{H}_2$, LiH\end{tabular} & Superconducting & \begin{tabular}[c]{@{}c@{}}Hardware-efficient\\ VQE\end{tabular} \\ \hline
		OTI Lumionics \cite{Ryabinkin-JCTC-2018} & 2018 & 4 & $\text{H}_2$, LiH & Superconducting & Qubit CC \\ \hline
		Li et al. \cite{PhysRevLett.122.090504} & 2019 & 2 & $\text{H}_2\text{O}$ & NMR & QPE \\ \hline
		IonQ/JQI \cite{Nam2020} & 2019 & 4 & $\text{H}_2\text{O}$ & Trapped-ion & VQE-UCC \\ \hline
		Oak Ridge \cite{McCaskey2019} & 2019 & 4 & NaH, RbH, KH & Superconducting & \begin{tabular}[c]{@{}c@{}}Hardware-efficient\\ VQE(-UCC)\end{tabular} \\ \hline
		Mitsubishi/IBM \cite{gao2019computational} & 2019 & 2 & Lithium superoxide dimer & Superconducting & VQE-UCC \\ \hline
		Smart \& Mazziotti \cite{PhysRevA.100.022517} & 2019 & 3 & $\text{H}_3$ & Superconducting & VQE-UCC \\ \hline
		Google \cite{arute2020hartreefock} & 2020 & 12 & \begin{tabular}[c]{@{}c@{}}$\text{H}_{6}$, $\text{H}_{8}$, $\text{H}_{10}$, $\text{H}_{12}$\\ HNNH\end{tabular} & Superconducting & VQE-HF \\ \hline
		IBM \cite{gao2020applications} & 2020 & 2 & \begin{tabular}[c]{@{}c@{}}PSPCz, 2F-PSPCz, \\ 4F-PSPCz\end{tabular} & Superconducting & \begin{tabular}[c]{@{}c@{}}qEOM-VQE\\ VQD\end{tabular} \\ \hline
	\end{tabular}
	\caption{Experiments with quantum computing hardware applied to simulating molecular- and material chemistry (this selection covers many, but not all results). Years and top-to-bottom order, in order of appearance on pre-publication service arXiv. Maximum qubit number denotes the number of qubits actually used in the simulation, potentially on sub-lattices of larger chips. All systems were discretized into near-minimal basis sets, i.e. STO-3G or similar, or utilized severe approximations to reduce the number of qubits in other ways. The ``superconducting'' platform denotes any variant of superconducting platform where microwave pulses are used to control qubits defined by flux or charge quanta on superconducting islands. VQE-UCC stands for any strategy combining the Variational Quantum Eigensolver algorithm with the chemistry-inspired Unitary Coupled Cluster ansatz approach. VQE-HF performs the Hartree-Fock procedure on-chip using VQE. IPEA and QPE are forms of quantum phase estimation implementations. The Quantum Equation-of-Motion VQE (qEOM-VQE) \cite{ollitrault2019quantum} and Variational Quantum Deflation (VQD) \cite{Higgott2019variationalquantum} methods are used to compute excited state energies.}
	\label{nisq_experiments}
\end{table*}

\section{Defining quantum advantage \label{sec:irrelevance}}

By quantum advantage one normally means a solution of a certain computational problem using a quantum computer that would be impossible in a reasonable time using any classical computer, including supercomputers. In this work we will restrict the discussion of quantum advantages to molecular chemistry problems only. Quantum advantage in molecular chemistry can be sought in a space of three dimensions: speed, accuracy, and molecule size. In order to demonstrate quantum advantage quantum computers must prove to be significantly more proficient than classical computers at handling any of these three dimensions. 

If achieved in application to any molecular problem, quantum advantage would be an impressive accomplishment on its own, but we must be aware that not every quantum advantage will be useful in practical or industrial applications. In order to seek transformative quantum advantages, we first need to indicate the types of advantages that will be of little value in practice.

\begin{enumerate}
	\item \textbf{Irrelevance due to availability of accurate experimental results}. The first type of irrelevant quantum advantage is where quantum computers would have to compete with experimental measurements that are accurate, fast, inexpensive, and straightforward. Some types of simulations rival experiment in how accurately they can probe reality (see examples of brilliant theoretical predictions in Refs.~\ocite{Schaefer-S-1986, Kroto-FST-1994}), but in presence of readily available, reliable experimental data there is little need for simulated results.
	
	\item  \textbf{Irrelevance due to availability of conventional computational results}. The second type of irrelevant quantum advantages pertains to chemical systems or problems for which quantum chemical calculations on classical computers can produce chemical accuracy results in little time -- seconds, minutes, or even a few hours. These are exactly the types of applications on which quantum computing algorithms have been routinely validated so far. They target gas phase energetics of small (diatomic or triatomic) molecules.\cite{Lanyon-NC-2010, Ryabinkin-JCTC-2018, Sugisaki-ACSCS-2019, Cao2019QuantumComputing} Reasonably fast conventional ab initio quantum chemical calculations are in their turn in competition with semi-empirical,\cite{Bannwarth-JCTC-2019} force field,\cite{Roos-JCTC-2019}, machine learning\cite{Smith2019} and composite\cite{Grimme-JCP-2015} solutions, which, as research progresses, converge toward chemical accuracy. Such economical approaches would also challenge quantum computers and weaken the value of any quantum advantage that may be achieved. For some types of property predictions, there is even no need for direct involvement of quantum chemical methods. In the pharmaceutical industry, for instance, binding affinities and solubilities can be predicted with adequate accuracy by molecular dynamics approaches which are based on force fields.\cite{Abel-ACR-2017,Mondal2019}
	
	\item \textbf{Irrelevance due to real world complexity}: One can imagine a quantum advantage to be eclipsed by the vast chemical and conformational complexity that often underlies real world chemistry. Often the biggest problem in simulation research is not simply to complete single computations in reasonable time with sufficient accuracy. When simulated chemical processes are very complicated and involve potentially hundreds of intermediates, conformations, or reaction paths, as in catalytic and metabolic pathways, the real research bottleneck lies in a
	combinatorial explosion of possibilities to probe with simulation. In such projects, even if the calculations themselves became orders of magnitude faster or more accurate (for example, through the exploitation of quantum computing), the whole project might enjoy only a modest speedup. A common way to reduce real world complexity of physico-chemical processes involving conformational, solvation, and thermal effects is the use of empirical methods. In practice, for example in the prediction of ADME/Tox properties in the pharmaceutical industry, such heuristics are often accurate enough to drive the discovery process.\cite{Lombardo-JMC-2017} 
	
	\item \textbf{Irrelevance to industrial applications}. Finally, there are molecular systems that do not fall under any of the three irrelevance categories exposed above, and yet they may be still irrelevant to quantum computing because they lack a direct connection with industrial applications. As such, these systems, even if described accurately on a quantum computer, are likely to remain academic curiosities and fail to lead to transformative changes in chemistry. 
\end{enumerate}

\section{Current capabilities of classical computers}
One way to search for quantum advantage is to establish the limits of computational power of classical computers, which will set the bar for quantum computers. If that bar presents a trivial problem for a quantum algorithm, we have found a quantum advantage and we will only have to check if the advantage is not \textit{irrelevant} (vide supra).

A good starting point for investigating the limits of quantum chemistry on conventional computers is to first assume that we are targeting chemical accuracy while using conventional basis sets and to establish the bar for systematically improvable quantum chemical methods such as configuration interaction\cite{sherrill-ci-1999} (CI) and coupled cluster\cite{crawford-cc-2000} (CC).

The largest conventional CI calculations which, to our knowledge, have been reported in the literature are proof-of-principle studies.\cite{Vogiatzis-JCP-2017} They include:  (i) a calculation involving a complete active space (CAS) with 20 electrons in 20 spatial orbitals, realized within the framework of the MCSCF method on a chromium trimer (corresponding to approximately $4.2\cdot10^{9}$ single determinants (SDs)); (ii) a single point CASSCF calculation on a pentacene molecule with 22 electrons in 22 spatial orbitals (corresponding to approximately $5.0\cdot10^{11}$ SDs); (iii) a single iteration of the iterative CI algorithm for a chromium tetramer with 24 electrons in 24 spatial orbitals ($\sim 7.3\cdot10^{12}$ SDs).\cite{Jeong-JCTC-2020} All these CI calculations utilized the 6-31G* basis set. To put the number of the SDs in chemical context, a FCI calculation on the propene molecule in the minimal STO-3G and the larger but still very small 6-31G basis set would correspond to 24 electrons in 21 and 39 spatial orbitals, respectively. The frequently used frozen core (FC) approximation would reduce the number of ``active'' electrons in propene to 18, but already the next homolog, 2-butene, will have 24 active electrons in the FC approximation, and push against the limits of the feasible FCI calculations. 

Among the largest CC calculations reported in the literature is a CCSDT(Q)/cc-pVTZ single point energy calculation on benzene in the FC approximation (30 electrons, 264 basis functions) which corresponds to ${\sim3.1\cdot10^{9}}$ single, double, and triple, as well as ${\sim 2.2\cdot10^{12}}$ perturbative quadruple t-amplitudes.\cite{Sylvetsky-JCP-2016} The computational scaling of CCSDT(Q) is very steep, namely the ninth power of the system size, but chemical accuracy can be achieved with significantly less expensive computational approaches. There is a multitude of composite methods that target chemical accuracy,\cite{Peterson-TCA-2012} but perhaps the best known, simplest, and most popular approach that satisfies many demands of the computational chemist for accuracy, including chemical accuracy for multiple types of systems, is the CCSD(T) method combined with a complete basis set (CBS) extrapolation. The CCSD(T) method is known to treat dynamic correlation accurately, but is generally inapplicable to molecular problems dominated by non-dynamic correlation. For more information on dynamic and non-dynamic correlation see Ref.~\ocite{Mok-JPC-1996}.

Likely, the largest conventional CCSD(T) calculation ever performed is that by Yoo and co-workers on a (H$_2$O)$_{17}$ cluster in the aug-cc-pVTZ basis set,\cite{Yoo-JPCL-2010} which corresponds to 128 electrons in the FC approximation and 1564 orbitals. A similar (H$_2$O)$_{16}$ calculation had to be executed on ${120000}$ computer cores and took more than 3 hours in year 2010. Another notable CCSD(T) effort is that of Gyevi-Nagy and co-workers reported in 2020,\cite{Gyevi-Nagy-JCTC-2020} in which the energy of 2-aminobenzophenone (ABP, C$_{13}$H$_{11}$NO) was computed in the large def2-QZVPPD basis set. That calculation, utilizing the density-fitting approximation, correlated 90 non-FC electrons among 1569 orbitals and was completed on 224 computer cores in 68 hours. The size of ABP makes it an attractive ``minimum viable'' system in ``real world'' practical applications. Systems of this size are regularly used for parameterizing force fields in areas of research like computational drug design.\cite{Roos-JCTC-2019} A recent work by Kruse and co-workers\cite{Kruse-JCTC-2019} reports sub-chemical accuracy, large basis set calculations using CCSD(T) on stacked DNA base pairs, which are even larger molecular systems.

Despite the general impracticality of the largest calculations mentioned above, just the sheer fact that they have been accomplished on classical computers sets the bar very high for quantum computers. Here we aim to define this bar in precise terms. For this, it will be convenient to assume that CCSD(T) with a large basis set is still the most trustworthy quantum chemical energy prediction method (at least for certain types of molecular systems), and neglect the small errors introduced by the density-fitting approximation, as in an application mentioned earlier in the text. In order to show a relevant quantum advantage in energy calculations targeting chemical accuracy, quantum computers must be able to dramatically outperform classical computers on a solution of, let us say for certainty, a 90 electron, non-relativistic electronic structure problem for an organic molecule at its near-equilibrium geometry. Conventional Gaussian basis set expansions, which have been so far considered for quantum computers, require about 1570 orbitals for such a calculation. A basis set of such a size appears insurmountable for the earliest available quantum computers. 

Conventional Gaussian basis set expansions provide a very slow convergence to the exact energy with the size of the basis set.\cite{Kong-CR-2012} Chemical and sub-chemical accuracy is achievable with explicitly correlated basis sets.\cite{Kong-CR-2012} Combined with density fitting techniques, they deliver an accurate result at a fraction of the cost of CCSD(T)/CBS.\cite{Sirianni-JCTC-2017, Ma-JCTC-2019} Recent  works\cite{motta2020quantum, mcardle2020improving} have considered the use of explicitly correlated F12 methods in near-term quantum computing applications and noted that these methods may lead to using less quantum resources. Given the potential reduction of required resources it seems like a more realistic approach than targeting chemical accuracy with conventional basis sets. 

Few researchers are eager to perform conventional, large-scale quantum chemical computations on classical computers due to their prohibitive cost. Recently there appeared CI- and CC-like methods that strive to approximate the results of very large and therefore unfeasible, proper CI and CC calculations, at low cost. For the CI-like methods see the recent report~\cite{Eriksen-A-2020} on the ground state of benzene and references therein. Among the CC-like methods especially noticeable is the domain-based local pair natural orbital (DLPNO) method.\cite{Liakos-JCTC-2015, Liakos2020} These new approaches achieve chemical accuracy on small organic molecules such as 1,3,5-hexatriene (for incremental FCI\cite{Zimmerman-JCP-2017}) or even small proteins (for LNO-CCSD(T))\cite{Nagy2019} within hours or days of computational time. Still other attractive approximate methods striving at chemical accuracy include quantum Monte-Carlo\cite{Booth-JCP-2009, Cleland-JCP-2010, Booth-JCP-2010}, some recent DFT functionals\cite{Mardirossian-JCP-2016} and neural networks.\cite{Smith2019} 

At first glance, it seems that the methods mentioned in the previous paragraph render quantum advantage that might be achieved on computation of energies of small, medium-sized, and even large organic molecules irrelevant. However, the applicability and accuracy of these approximate methods becomes less certain vis-\`{a}-vis systems with a strong multireference character (non-equilibrium geometries, radicals, presence of transition metal atoms), apart from other limitations.

\section{Comment on quantum advantage in FeMo-co research\label{sec:FeMo-co}}

Multiple publications intended for the general audience as well as several research publications\cite{Reiher2017ElucidatingComputers, Li-JCP-2019} reflect on the potential application of quantum computers to the particularly challenging and, to some degree, mysterious natural catalytic complex known as FeMo-co. The core of this catalytic protein system, which reduces the atmospheric N$_2$ to biologically processable forms of nitrogen, is comprised of eight transition metal atoms (seven iron atoms and one molybdenum atom) interlinked by sulfur and carbon atoms. Numerous experimental\cite{Milton-ACR-2019} and theoretical\cite{Thorhallsson-CS-2019} studies have been attempting to decipher the catalytic mechanism through which FeMo-co utilizes the nearly inert N$_2$. So far, classical computers were largely incapable of treating this system with accuracy and computational efficiency necessary for cracking the mechanism.\cite{Foster2018, Dance-I-2019} 

Being a system that contains multiple transition metal atoms, FeMo-co is expected to present a strong correlation problem, and therefore both dynamic and non-dynamic correlation\cite{Mok-JPC-1996} need to be recovered for its accurate solution. The minimal, 39-atom model of one of its protonation states has the stoichiometric formula [C$_{7}$H$_{9}$Fe$_7$MoN$_2$O$_3$S$_{10}]^{-3}$, which yields 254 active electrons in the FC approximation. What is the minimal (N,N) CAS needed to describe the non-dynamic correlation in this system with chemical accuracy?

In year 2017 Reiher and co-workers\cite{Reiher2017ElucidatingComputers} estimated that a CAS of the size (54, 54), which is far larger than what CASCI or CASSCF can address on a classical computer, should be within our computational means to treat on a quantum computer. Would such a CAS be sufficient for an accurate, converged description of non-dynamic correlation in the FeMo-co system? In 2018 Montgomery and Mazziotti actually conducted CASSCF and V2RDM calculations for a FeMo-co model with increasing, but still very small basis sets (STO-3G, 3-21G, DZP) and increasing CASs, with up to (30, 30) in case of V2RDM.\cite{Montgomery-JPCA-2018} The results of these authors indicate that their most accurate energy calculation at the V2RDM/DZP/(30, 30) level is nowhere near being converged. The difference in energies between the V2RDM/X/(26, 26) and V2RDM/X/(30, 30) levels of theory is 50.6 kcal/mol, 36.6 kcal/mol, and 106.1 kcal/mol for X = STO-3G, 3-21G, DZP, respectively. Not only is the energy far from being convergent with respect to the size of CAS but it is also far from being converged with respect to the size of the basis set. In 2018 Tubman and co-workers performed a (54, 54) CAS calculation with the approximate ASCI method applied to FeMo-co.\cite{tubman2018postponing} The convergence of the FeMo-co energy with the size of the CAS needs to be studied further, but given the gigantic energy variations with the CAS around (30, 30) seen in the work of Montgomery and Mazziotti it is unlikely that a (54, 54) CAS is close to being sufficient for convergence of non-dynamic correlation in the FeMo-co system. Indeed, a recent work by Li and co-workers estimates that a (54, 54) CAS for FeMo-co is expected to yield a qualitatively incorrect, single reference wave function.\cite{Li-JCP-2019} These authors propose a (113, 76) CAS for this system instead and use it in a preliminary DMRG\cite{Chan-ARPC-2011} calculation with bond dimension 2000. This calculation was followed by an analogous (113, 76) CAS but more accurate DMRG calculation with bond dimension 6000, which was completed on a supercomputer containing 2480 cores.\cite{brabec2020massively} 

Dynamic correlation effects would have to be accounted for separately. Reiher and co-workers propose to recover the missing dynamic correlation through a method like DFT on a classical computer.\cite{Reiher2017ElucidatingComputers} However, the accuracy of such a combined treatment would be quite uncertain due to uncontrolled errors of DFT. Levine and co-workers resort to corrections through the second-order Epstein–Nesbet perturbation theory (PT2).\cite{Levine-JCTC-2020} Their promising claim that PT2 gives ``extremely accurate results, often within a kcal/mol of the absolute FCI energy''\cite{Levine-JCTC-2020} must be, however, tested on a system of the size and complexity of FeMo-co. 

We must also not forget that the FeMo-co molecular system contains an atom of the element molybdenum, which is heavy enough to merit treatment with relativistic methods or pseudopotentials in studies of inorganic complexes\cite{Neyman-O-1997, Alemayehu-CC-2017, Weidman-JPCA-2018} and enzymes.\cite{Ryde-2017} In particular, relativistic calculations have been already used in FeMo-co research.\cite{Bjornsson-CS-2014, Rees-DT-2017, Bjornsson-IC-2017} 

Extending non-relativistic quantum chemistry to a relativistic variant, on quantum computers, has been investigated in Refs.~\ocite{veis2012quantum1, veis2012quantum2} where it was shown that, in the direct mapping in second quantization, the scaling of the relativistic form of the non-relativistic counterpart algorithm was identical, assuming the no-pair approximation. Relativistic calculations on classical computers have been shown to require large basis sets in combination with extrapolation schemes for convergent energies, not unlike in non-relativistic calculations.\cite{Peterson-JCP-2007} 

Even if van der Waals interactions and other types of important effects, no doubt imposed by the protein matrix, could be ignored, to properly model a minimal size, 254-electron, strongly correlated system, the quantum computer algorithm would have to show a quantum advantage on what would be equivalent to a multireference, relativistic calculation with a CAS being presumably much larger than (54, 54). We believe that Reiher and co-workers' attempt to work with the CAS of this size\cite{Reiher2017ElucidatingComputers} was a step in the right direction, but it is likely to prove only the first step. In future research on attempting to predict the electronic structure of FeMo-co on quantum computers it will be important to incorporate knowledge about the convergence of the non-dynamic correlation energy with respect to the size of the active space. We also need to make sure that we can recover the remaining, dynamic correlation accurately.  

An attempt to accurately compute the dynamic correlation of the minimal FeMo-co model with a quantum computing algorithm runs into the problem of the large basis set, and, as a consequence, into the present unavailability of the equivalent number of logical, coherent qubits. For guaranteed chemical accuracy we have to assume the need for at least a triple-$zeta$ basis set. Such a minimal basis set might be cc-pVTZ applied to all atoms except Mo (where it is appropriate to use the pseudopotential cc-pVTZ-CC), bringing the total number of spatial orbitals to 1365, or better cc-pVQZ and cc-pVQZ-CC, respectively, resulting in 2367 orbitals. In order to reduce the number of basis functions, an F12-adapted basis set such as def2-TZVPP on the metals and def2-TZVP on the lighter elements, as recommended by Kesharwani and Martin,\cite{Kesharwani-TCA-2014} could be used, resulting in 1398 basis functions. So, dynamic correlation, recoverable with systematically improvable accuracy, presents a problem for quantum computers because it requires basis sets that are too large for any quantum computer in the near future. Therefore, there is need for more research into constructing and validating methods that would combine non-dynamic correlation obtainable on a quantum computer with dynamic correlation obtainable with methods like DFT or PT2 on a classical computer. 

\section{Quantum computational chemistry}
\subsection{Potential of quantum computational chemistry: an introduction}
A quantum computer can in principle prepare and store richer many-body wave function representations than classical computers.  There are highly efficient numerically exact quantum-computational algorithms,\cite{Nielsen2010QuantumInformation} executable in polynomial time, for simulating time evolution under both  time-independent and time-dependent many-body Hamiltonians such as those used in quantum chemistry models. Conversely, in classical computers the numerically exact diagonalization and CI methods operate in the space of determinants of large matrices which requires an exponentially scaling number of operations.

At first glance, these fundamental differences give some indication that quantum computers may be perfectly suited for simulating chemistry. There are, however, some caveats and significant challenges when implementing quantum computational methods in practice; in \tabref{nisq_experiments} we show progress in quantum computational chemistry experiments in recent years. In the following sections we detail some of the challenges in the short and long term specifically for molecular quantum chemistry. We refer the interested reader to review papers\cite{Whitfield2011d,McArdle2018h,Cao2019QuantumComputing} for more details. Next, we consider several examples of end-to-end quantum resource and time estimates for calculating ground state energies with a CAS approach on a theoretical fault-tolerant quantum computer (FTQC).

Broadly speaking, there are two widely discussed regimes in the quantum computing community, ``near term'' and ``long term'' methods and hardware. We stress that it is currently uncertain in how many years from today, ``near term'' changes to ``long term''. Rather, in the context of quantum chemistry algorithms, these terms are often loosely distinguished by the different ways in which Hamiltonian expectation values are measured on the quantum computer, and whether fault-tolerant error correction is applied. In this work we focus on calculations of ground state energies, as excited states and dynamical properties can be accessed with similar techniques and with similar complexity arguments.

There are two main paradigms orchestrating quantum Hamiltonian simulation algorithms. In \textit{Hamiltonian averaging},\cite{Peruzzo2014} Hamiltonian expectation values are extracted by performing partial tomography on a prepared quantum state of the qubit register, which in turn models the chemical system's many-body quantum wave function. The number of repetitions (samples) in this tomography scales as $\mathcal{O}(1/\epsilon^2)$ to reach an energy estimation error $\epsilon$. For each repetition the same superposition state needs to be prepared again.
In \textit{Hamiltonian phase estimation}, an approximate eigenstate of a Hamiltonian is prepared once and a series of $\mathcal{O}(1/\epsilon)$ \textit{phase estimation} circuits are applied to the qubit register in order to estimate the eigenstate's energy to accuracy $\epsilon$.\cite{Kitaev1995, Wiebe2016}

In terms of scaling of the quantum computational runtime with accuracy, the phase estimation method is clearly superior. However, in the current generation and near-term quantum computing devices, maintaining qubit coherence, performing measurements, gate operations and state preparations with high fidelity is still very challenging (see, for example, references from \tabref{nisq_experiments}). This does not allow for the phase estimation circuit with gate depth $\mathcal{O}(1/\epsilon)$ (with a large pre-factor) to be executed faithfully in the near term. Fault-tolerant circuit execution is required, and its practical implementation at the scale required for quantum advantage is not expected until longer term. In the next section we use Hamiltonian averaging for illustrating the potential of quantum computational chemistry in the short term as well as its associated challenges. 

\subsection{Potential of quantum computational chemistry: short term}
One popular algorithm for simulating quantum chemistry on currently available quantum devices is the variational quantum eigensolver (VQE).\cite{Peruzzo2014} The algorithm works by first mapping the quantum chemistry model Hamiltonian to a qubit Hamiltonian. Subsequently, a trial quantum state is prepared using a so-called Ansatz circuit. Next, one may measure any operator expectation values, such as the Hamiltonian energy, over this state. A variational optimization of this energy may converge to a good approximation of the ground state energy as long as the Ansatz circuit has sufficient expressability and the optimizer finds a good local or global minimum. The measurement circuit, and therefore the overall circuit, is relatively short, which is favourable to near-term noisy intermediate-scale quantum (NISQ) devices.\cite{Preskill2018c} Additionally, the variational aspect may allow for partial compensation of gate errors. Fundamentally, for perfectly noise-free execution, the accuracy of the VQE method is limited by the expressability of the Ansatz circuit. This aspect is similar to the limitations of classical (truncated) coupled cluster methods, or any methods which do not include all possible electronic excitations, or all configuration state functions, as FCI does. 

It is difficult to estimate an exact runtime of the VQE method due to its variational nature.\cite{Kuhn2018AccuracyComputer} However, a ballpark estimate of the runtime of each iteration step may be calculated, based on the details of the algorithm and estimates of hardware parameters such as quantum gate times and measurement and reset time. This allows one to at least perform a sanity-check on the feasibility with respect to the runtime for a given application.

With such an estimation we investigate the feasibility of NISQ-VQE to simulate specifically dynamical correlations of molecular systems, which are just in reach for simulation on classical (high-performance) computers. In the ABP molecule discussed earlier, 1569 orbitals and 90 non-frozen core electrons were considered, a calculation which took 68 hours to complete on a large computer cluster. We estimate that a state-of-the-art quantum algorithm strategy like k-UpCCGSD \cite{Huggins2018} in a VQE approach for the same situation with chemical precision in the same basis would require over 3000 physical qubits and take completely prohibitive time, i.e. centuries per single iteration of the optimizer, even with severe approximations like the paired-electron assumption, pUCCD.\cite{Elfving2020SimulatingDevice} We use hardware parameters typical to superconducting qubit devices and ignore the overhead of connectivity. We also assume full circuit parallelizability and zero latencies. 

It is still unknown how NISQ-VQE approaches perform for simulating molecular systems dominated by non-dynamic correlations. If we assume that one could reach CASCI/CASSCF levels of precision in a given basis set, by using a NISQ-VQE methods with a mere linear circuit depth, one can estimate with similar assumptions as above, that tackling a (26,26) CAS problem would take about 1 hour \textit{per VQE iteration}. Although this runtime does not seem completely prohibitive, the range of assumptions made, prevents us from drawing strong conclusions for practical and industrially relevant application. In second quantization, the number of electrons scales linearly with the basis set size, N, and therefore the number of two-qubit gates for any relevant ansatz should be at least of size N, resulting in at least a depth-N circuit when nearest-neighbour qubit connectivity is combined with a swap-network approach.\cite{PhysRevLett.120.110501}. Using first quantization methods could improve the scaling in time and space in terms of the basis set size, but likely still require larger logical qubit counts for reaching chemical accuracy.\cite{Babbush2019sublinear} Regardless of the specifics of currently-known methods, either orders of magnitude improvements in gate, control and measurement speed may be required or massive clusters of distinct quantum computers working in parallel,\cite{vanMeter2006,distributed2013,modular2014} to keep the runtimes practical.

Besides a solution to the runtime complications, major hardware and algorithmic progress is required in order for NISQ-VQE to attain these theoretically predicted accuracies; even the relatively shallow quantum circuits in VQE implicate significant impact of (in)coherent noise processes on the estimation of energy accuracies.\cite{franca2020limitations} Partial error mitigation techniques exist, but all incur significant additional quantum or classical/runtime overheads, exacerbating the runtime challenge.\cite{McArdle2018h}

We stress that above estimations and challenges of NISQ-VQE were only considered in the context of molecular chemistry simulations with mostly dynamical correlations, which may potentially turn out to be relatively less promising as compared to other applications. Outside the focus of this work, there are many strategies for applying NISQ-compatible algorithms to simulating material chemistry, solid-state physics, field theories and more.\cite{Babbush2018l, Ma2020, tamiya2020calculating, hubbardEstimation, yao2020gutzwiller, kim2017robust, Kokail2020, PhysRevA.98.032331} It is important to estimate the resource requirements, including pre-factors, for their respective applications and gauge them against the corresponding classical computational limits.

\subsection{Potential of quantum computational chemistry: longer term}
Longer-term, fault-tolerant quantum computers\cite{aharonov1996fault} promise to solve both the problems associated with noise and the runtime issues. Error-correction schemes\cite{Roffe2019b} suppress circuit errors, allowing faithful execution of quantum computational chemistry algorithms at practically unlimited circuit depth. Higher individual gate error rates imply larger overheads in terms of physical qubits per logical qubit. Also, larger error correcting codes have longer clock-cycle times, which increases overall runtimes. It is therefore only feasible to construct large-scale fault-tolerant architectures when hardware error rates are low enough to implement reasonably sized codes while adhering to the error threshold.\cite{aharonov1996fault}

Fault-tolerant execution unlocks the use of the deep circuits commonly associated with quantum phase estimation (QPE).\cite{Kitaev1995,Abrams1997,Abrams1998} In QPE for quantum chemistry, Hamiltonian phase estimation is used to extract the eigenenergies associated with prepared eigenstates of a Hamiltonian. Besides the favourable \textit{scaling} with energy estimation accuracy $\epsilon$ (runtime scaling with $\mathcal{O}(1/\epsilon)$ and polynomial in system size),\cite{Kitaev1995, Wiebe2016} the energy can be found to \textit{arbitrary accuracy} (within the basis set used), in contrast to the VQE technique. This is because the measurement collapses the prepared approximate eigenstate to an exact eigenstate of the Hamiltonian, with probability equal to the overlap between these states (which can be achieved with at most a polynomial depth ansatz circuit)\cite{tubman2018postponing}. Its associated eigenenergy is then found, to a precision equal to the number of measurement bits used. This makes QPE also suitable for simulating non-dynamic correlation, which is extremely advantageous in view of the steep scaling of classical CI techniques.

Viewing these observations as promising, we move on to calculate the runtime and resource estimates of the QPE algorithm for relevant applications in chemistry. This includes a calculation of error-correction overhead which we also perform in the following sections. For similar calculation methods see Refs.~\ocite{wecker2014, Reiher2017ElucidatingComputers, Kivlichan2019, burg2020quantum}. 

\subsection{Resource estimates: Number of gates}
For the resource estimates, we consider two applications, the hydrogen molecule and the chromium dimer. We consider the resource estimates for simulating the ground state energy of the hydrogen molecule, $\text{H}_2$, to chemical accuracy at equilibrium bond length, $0.707$ \r{A}. This is a system of just 2 electrons which possesses mostly dynamical correlation. In order to reach true chemical accuracy (and not just precision), one would need to make an extrapolation over a series of basis sets of increasing sizes in order to approximate the complete basis set. The reason we focus on this very basic example is because so far most quantum computational research has focused on simulating the hydrogen molecule in tiny basis sets (STO-3G or 6-31G, see \tabref{nisq_experiments}), and claimed chemical accuracy where  chemical precision within a limited basis set should have been referred to. Here we show expected resource estimates for obtaining chemical accuracy with respect to experimentally obtained values. We do this by taking a progressively larger basis set from the list $\{$STO-3G, 3-21G, 6-31G, cc-pVTZ, cc-pVQZ$\}$, which corresponds to using {2, 4, 10, 28, 60} spatial orbitals, respectively. 

We also consider the ground state energy of the chromium dimer, $\text{Cr}_2$, at its equilibrium bond distance, $1.68$ \r{A}. This molecule, with its very short, formally sextuple bond, and a peculiarly shaped dissociation curve, has been viewed as a critical test for electronic structure methods.\cite{Brynda-CPL-2009, Vancoillie-JCTC-2016, chromium_classical} Our focus is not specifically the single point energy or the potential energy curve but the potential of quantum algorithms for studying the chromium dimer at a higher accuracy than has been possible with classical computers algorithms.

For the study of the significant \textit{non-dynamic} correlation of the chromium dimer we choose the complete active space (CAS) approach, where we select an increasing number of active orbitals and active electrons. The orbitals were computed with the RHF/cc-pVTZ level of theory and the frozen core approximation was not activated. In what follows we perform resource estimation on ground state energy estimation within the chosen active space on the quantum computer. In this way, we can compare the obtained energies with classical CASCI energies. For further comparison, we conducted CASSCF calculations and used the final rotated basis to construct the input second-quantized Hamiltonian to the quantum oracle constructor. This typically yields a denser (less sparse) Hamiltonian. This is a good indicator of the final CI steps of a CASSCF calculation when performed on a quantum computer, as opposed to the initial CI steps starting from a single-reference like (R)HF. 
\begin{figure}[ht]
	\begin{tabular}{l}
		\hspace{1pt}\includegraphics[width=0.91\linewidth]{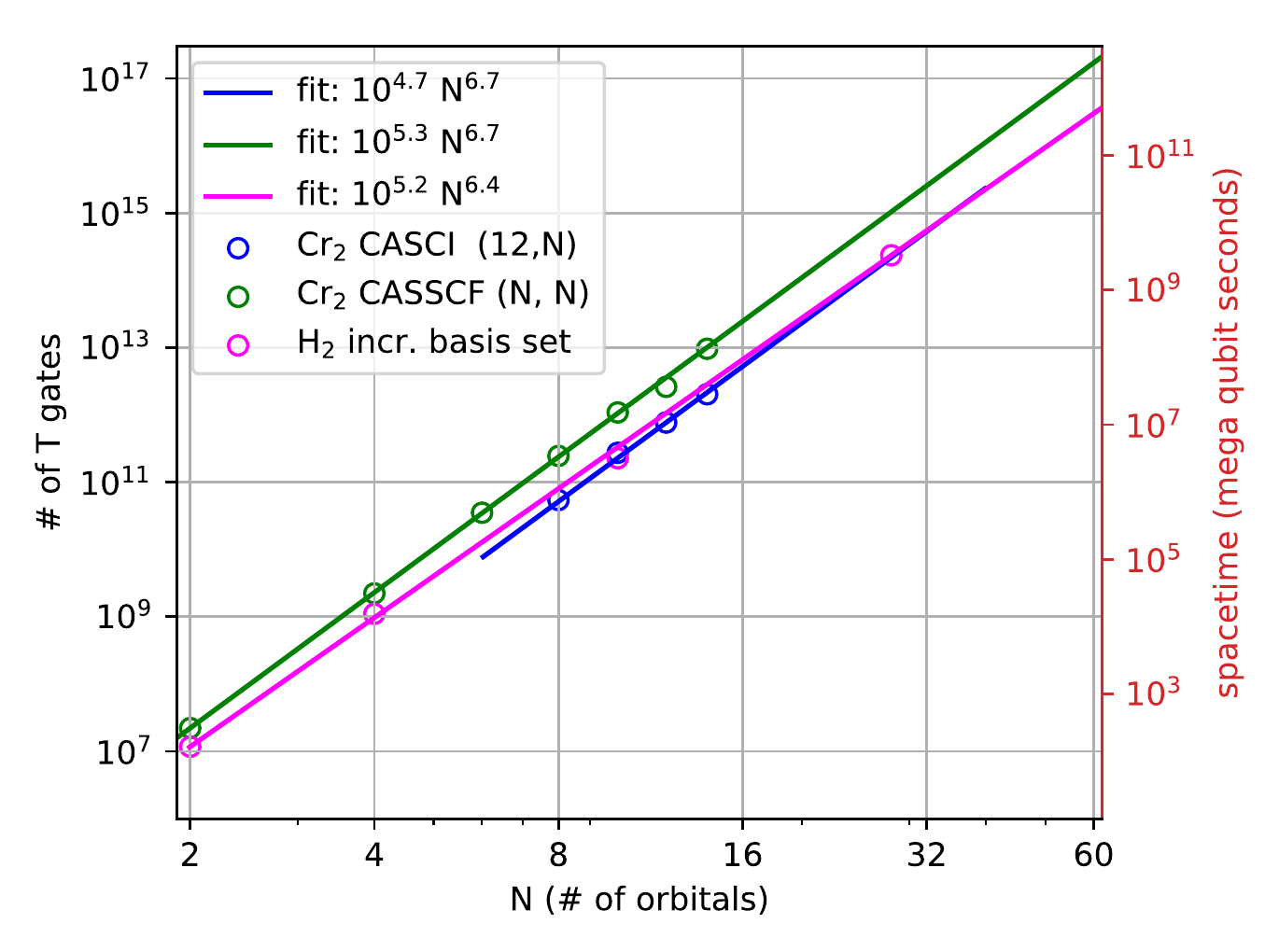}\\
		\includegraphics[width=0.9\linewidth]{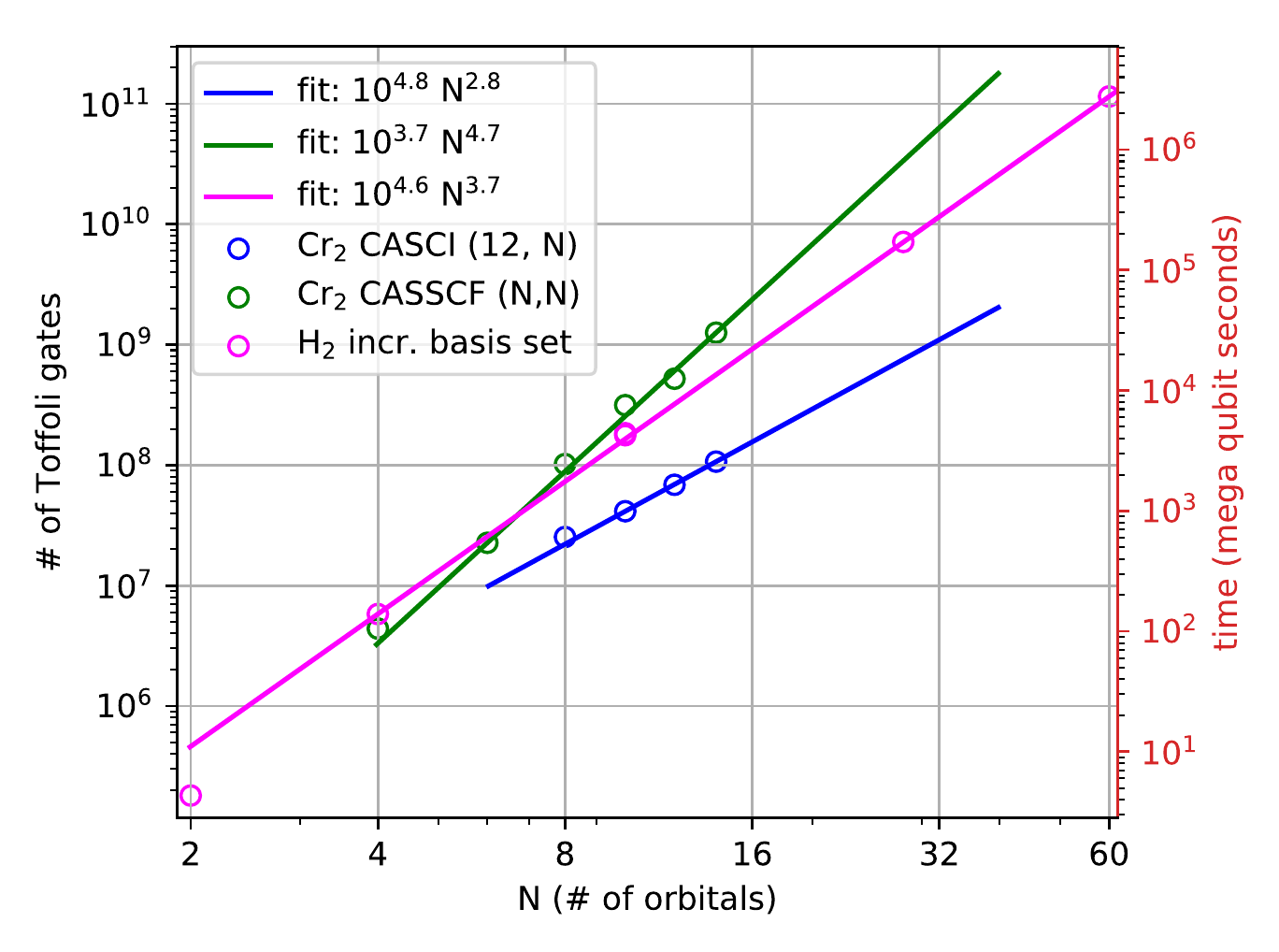} \\
		\hspace{4pt}\includegraphics[width=0.8\linewidth]{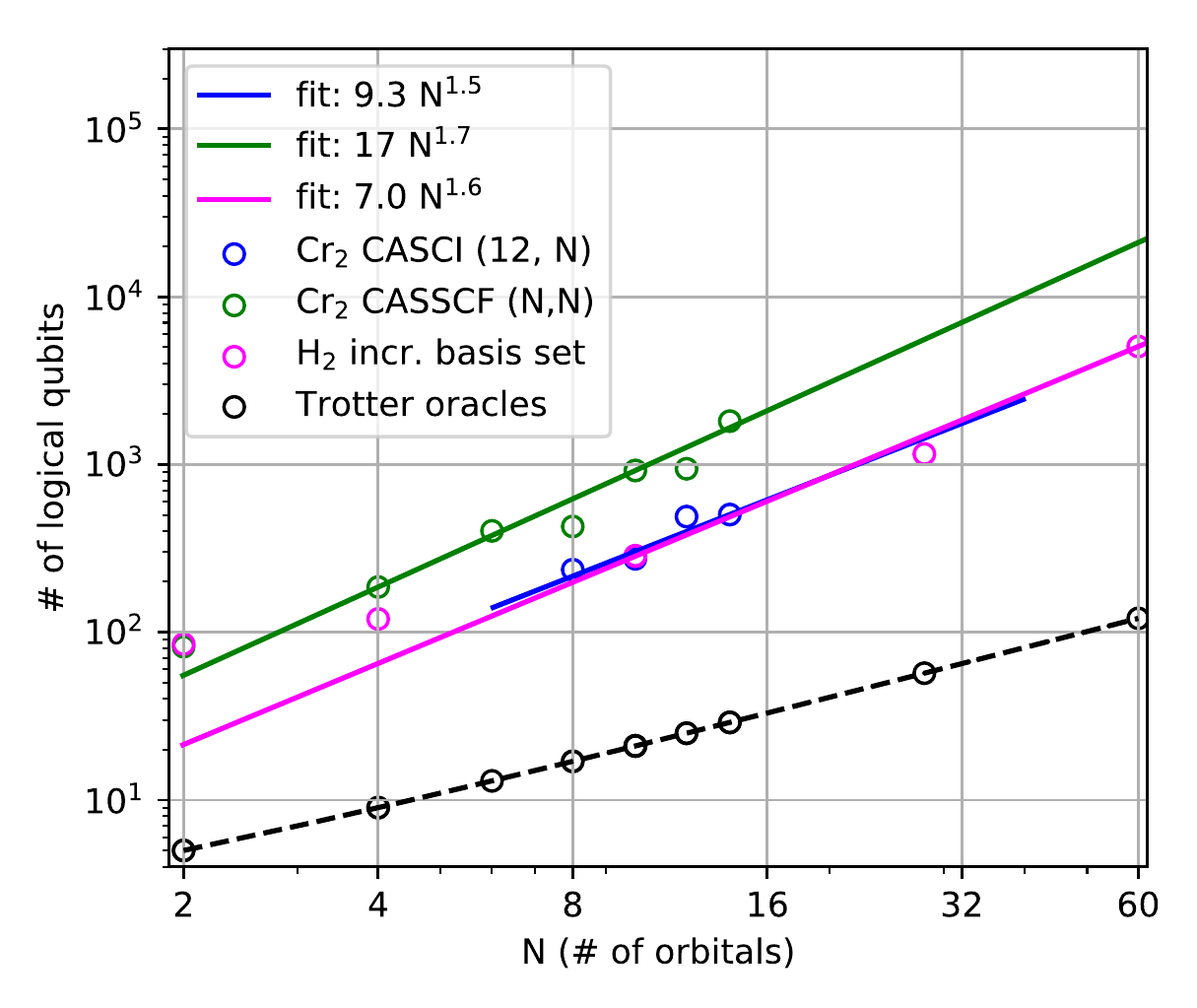} 
	\end{tabular}
	\caption{(Top) number of T gates for a Trotterization implementation. (Center)  number of Toffoli gates for a qubitization implementation. We assume a spacetime complexity of 24 qubitseconds per-Toffoli gate, and 14 qubitseconds per -T gate using a `C2T' factory, both from Ref.~\ocite{Gidney2019efficientmagicstate} (Bottom) The total number of logical qubits for the qubitization and Trotterization protocols used here.\label{scaling_qpe_resources}}
\end{figure}
There are many different schemes for the Hamiltonian-simulation part of QPE; two important classes are Trotterization and qubitization, which we consider here. 

\subsubsection{Trotterization}\label{ResEsTrZ}
In this section we provide resource estimates for the iterative QPE algorithm with oracles based on a Trotter-Suzuki decomposition\cite{Suzuki1993} of the direct Hamiltonian simulation operator $\hat{U}=e^{-i\hat{\mathcal{H}}t}$. We aim to find the optimal resource estimates for ground state energy simulation to chemical precision within the chosen basis set. As we consider the class of iterative phase estimation techniques, a single ancilla logical qubit is added to the $2N$ spin-orbital-representing qubits for a total of $2N+1$ logical qubits. We are interested in minimizing the total T gate resource cost. This is also approximately equivalent to minimizing the wall-clock runtime due to the dominating fault-tolerant implementation cost of T gates in popular fault-tolerant implementations.\cite{fowler2012} We use the same derivation for the Trotter resource costs as in Ref.~\ocite{Reiher2017ElucidatingComputers}. Here, for simplicity, for the Trotter number we assume the same shifted scaling in terms of the system size (in terms of the number of orbitals) as in Fig. 4 of the Supplementary material of Ref.~\ocite{Reiher2017ElucidatingComputers}. We strongly emphasize that this assumption is merely an approximation that should generally be investigated case-specifically to guarantee any resource estimation bounds.

In \figref{scaling_qpe_resources} we plot the T gate cost, for a simulation that reaches chemical precision within the respective basis set, as a function of the number of orbitals. We find as expected a clear polynomial scaling in the system size, to less than seventh order. Besides a CASSCF (N,N) calculation, we also consider a CASCI(12,N) case where the number of electrons are fixed to 12 and we adjust the number of spatial orbitals $N$.  We find very similar apparent scaling and pre-factors for all three cases considered, although this could be primarily due to the assumption of similar scaling in the required Trotter number. It is clear that the CASSCF-optimized basis set, compared at $N=12$, requires a larger number of gate operations than a CASCI-step which has a single-reference character. This is due to the larger number of non-zero terms in the Hamiltonian in the rotated basis, leading to a larger number of terms in the Trotter expansion. We can roughly estimate the spacetime complexity of the error-correction overhead and find that millions to billions of qubitseconds are required for such simulations.

\subsubsection{Qubitization}\label{ResEsQBZ}
In Ref.~\ocite{Berry2019qubitizationof}, an efficient molecular chemistry simulation method was proposed, combining elements of, and improving on, a large body of prior work.\cite{Low2016, Berry42,Berry44, Babbush2018EncodingComplexity} The paper discusses several different techniques, including one leveraging an efficient low-rank representation of the Coulomb operator. However, we found the sparse method to work the best in our examples, as well as for simulating the reference molecule FeMo-co. With the sparse method, there is a relatively large overhead in the number of logical qubits but it is optimized in terms of the non-clifford gate complexity. The algorithm relies mostly on Toffoli gates and therefore we express resources in terms of Toffolis. We refer the interested reader to Ref.~\ocite{Berry2019qubitizationof} for details on how one may calculate these resource estimates.

In \figref{scaling_qpe_resources} we plot the gate cost of the qubitization method as a function of number of orbitals. We see significant differences in scaling with respect to the number of spatial orbitals which are included, which indicates a greater difficulty in calculating CASSCF (N,N) energies for the chromium dimer than for hydrogen with basis set size N using qubitization. Both the scaling and pre-factors of the fitted curves are better than in the provided Trotterization example. The estimated spacetime complexity is reduced by about 4 orders of magnitude across the board.
In the bottom diagram from \figref{scaling_qpe_resources} we plot the number of logical qubits required for qubitization, and we may compare this to the number required by the Trotterization scheme. The latter is independent of the system specifics and depends only on the characteristic system size $N$ because of the direct spin-orbital to qubit mapping. The qubitization scheme requires the availability of additional logical qubits depending on the number of non-zero terms in the system Hamiltonian. The trade-off between Trotterization and qubitization is then the circuit depth versus number of logical qubits.

\subsection{Resource Estimates: Error-correction overhead}\label{ResEsECO}
The quantum algorithms discussed above have favourable scaling with the size of the chemical system (polynomial time for arbitrary accuracy energy estimation). However, there is still a significant pre-factor making the total gate count large even for threshold-advantage systems. Furthermore, an even greater overhead comes from the need for error correction. A large body of work covers competing fault-tolerant strategies existing today; here we implemented two particular state-of-the-art techniques for executing fault-tolerant non-Clifford gates on the surface code,\cite{fowler2012} detailed in Refs.~\ocite{Gidney2019efficientmagicstate, Litinski2019gameofsurfacecodes}.

Note that for the purpose of this resource estimation we have narrowed our hardware related assumptions to superconducting qubit quantum-processors with 2D nearest-neighbour qubit connectivity, which is a type of quantum-processor that is among those which became historically \textit{commercially} available first. However, superconducting qubits are only one of many different types of qubit architectures. Given that the current and expected operating parameters of quantum processors based on superconducting qubits, trapped ions, neutral atoms, photonics, quantum-dots, topological qubits, or other types, can differ significantly, it is not at all clear at this time which of these architectures will be first to enable such relevant quantum-advantage. In particular, the connectivity will greatly affect the error-correction overhead, whether that is nearest-neighbour, intermediate, or all-to-all connectivity.\cite{roffe2020decoding, webber2020, PhysRevA.89.022317} The interplay between a given connectivity-level and error-correction overhead is an active field of research.

\begin{figure}[ht]
	\begin{tabular}{l}
		\includegraphics[width=0.99\linewidth]{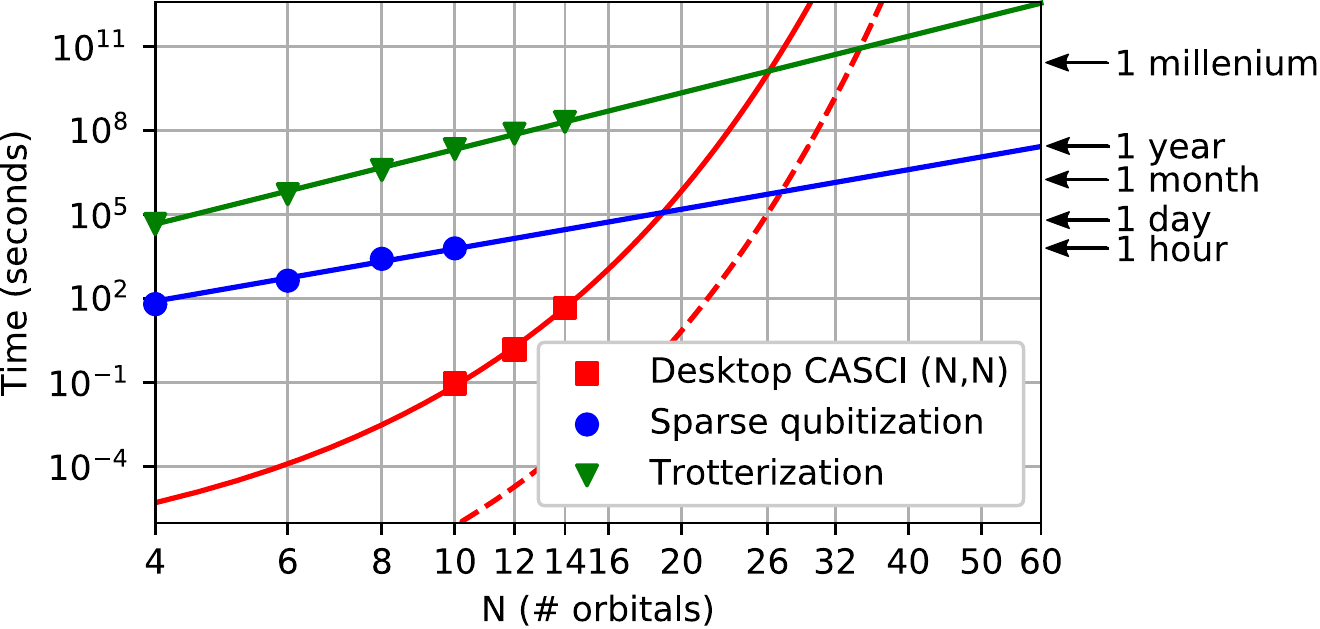}\\
		\includegraphics[width=0.82\linewidth]{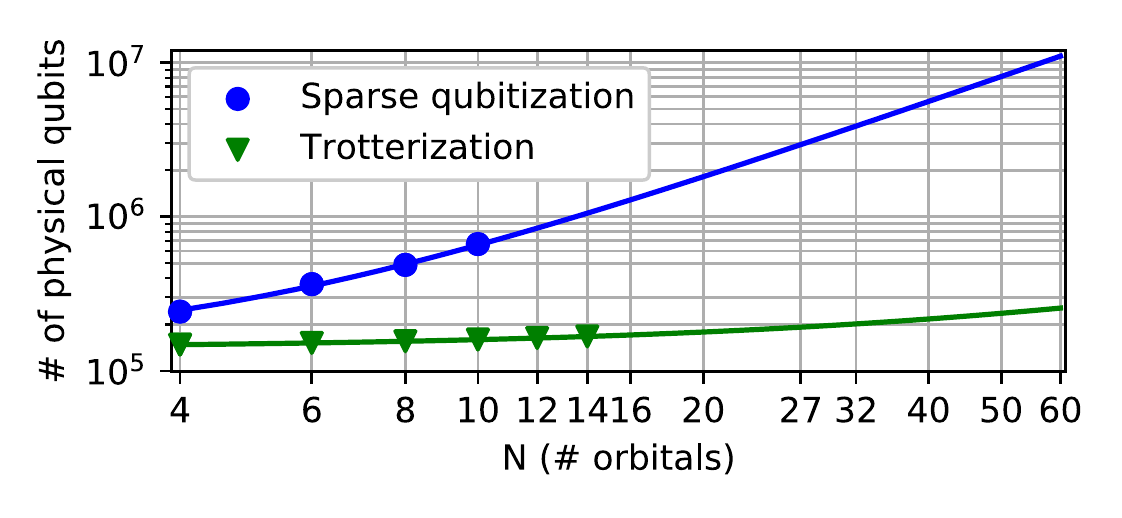}
	\end{tabular}
	\caption{Wallclock time scaling (top) and total physical qubit requirements (bottom) for the chromium dimer $\text{Cr}_2$ CASCI (N, N) simulation, for sparse qubitization and Trotterization algorithms running on a fault-tolerant quantum computer, optimized for less physical qubits, and comparing to a desktop PC simulation (full red line, corresponding to Intel i9-10980XE, with $\sim 1.2$ TFLOPS)\cite{intel} or a $10^5$x faster extrapolation (dashed red line, corresponding to a top-5 HPC, at $\sim 125$ PFLOPS)\cite{top500}. Curves represent curve-fitting while markers represent numerical instance-specific data. \label{scaling_wallclock}}
\end{figure}

In \figref{scaling_wallclock} we plot an example wallclock time estimate for performing a CASCI (N,N) simulation of the chromium dimer at equilibrium bond distance, using either a standard desktop PC, a Trotterization approach using a single Toffoli factory from Ref.~\ocite{Gidney2019efficientmagicstate}, or a qubitization approach using single distillation block from Ref.~\ocite{Litinski2019gameofsurfacecodes}.

The figure shows that the approximate size where a quantum computer solves the problem faster than a classical computer is for a (N,N) CAS size of around $N=19-34$. For $N>34$ any of the assessed quantum algorithms should be faster than any available classical computer. $N=19$ implies a physical qubit count of $\sim10^5$ for Trotterization and $\sim3\cdot10^6$ for sparse qubitization. However, in the case of Trotterization, the crossover point happens at a total runtime approaching a thousand years, which even with some optimization operations seems infeasible. The crossover for qubitization appears to happen at approximately the same point as the maximum size which is still feasible at all on a classical (super)computer. We stress here that the large number of assumptions which went into these calculations make it hard to pin the exact $N$ for the crossover.

\begin{table}[ht]
	\begin{tabular}{|c|c|c|c|}
		\hline
		& \begin{tabular}[c]{@{}c@{}}Optimized \\ for\end{tabular} & \begin{tabular}[c]{@{}c@{}}Number of \\ physical qubits\end{tabular} & \begin{tabular}[c]{@{}c@{}}Total \\ runtime\end{tabular} \\ \hline
		\multirow{2}{*}{\begin{tabular}[c]{@{}c@{}}Trotterization\\ $p=10^{-3}$\end{tabular}} & space & $3.8\times 10^5$ & 1485 years \\ \cline{2-4} 
		& time & $1.6\times 10^6$ & 161 years \\ \hline
		\multirow{2}{*}{\begin{tabular}[c]{@{}c@{}}Trotterization\\ $p=10^{-6}$\end{tabular}} & space & $2.0\times 10^4 $ & 343 years \\ \cline{2-4} 
		& time & $8.6\times 10^4$ & 37 years \\ \hline
		\multirow{2}{*}{\begin{tabular}[c]{@{}c@{}}Qubitization\\ $p=10^{-3}$\end{tabular}} & space &  $4.6\times 10^6$ & 43 days \\ \cline{2-4} 
		& time & $7.1\times 10^6$ & 110 hours \\ \hline
		\multirow{2}{*}{\begin{tabular}[c]{@{}c@{}}Qubitization\\ $p=10^{-6}$\end{tabular}} & space & $2.7\times 10^5$ & 11 days \\ \cline{2-4} 
		& time & $4.2\times 10^5$ & 27 hours \\ \hline
	\end{tabular}
	\caption{Total resource estimates for simulating the chromium dimer at equilibrium bond distance ($1.68$ \r{A}) with a CAS space of $(26,26)$ within a cc-pVTZ basis set. The number of logical qubits for the actual problem computation is 53 for the Trotterization strategy and 1366 for the qubitization strategy. We compare the results at two different levels of error rates, $p=10^{-3}$ and $p=10^{-6}$. The number of physical qubits includes both those required to store the data and those for the state distillation protocol.\label{detailed_results}}
\end{table}

Next, we make a more detailed estimate for a $\text{Cr}_2$ CASCI (26,26) calculation, as such a computation seems on the threshold of staying infeasible even for the coming years on classical supercomputers. We now also include the possibility of parallel distillation of non-Clifford gate operations; for this reason, we focus on the methods presented in Ref.~\ocite{Litinski2019gameofsurfacecodes} which takes this factor into account. Here we neglect effects of routing and memory buffers, and direct the reader to Ref.~\ocite{Litinski2019gameofsurfacecodes} for more details on that matter. 

\begin{table*}[ht]
	\includegraphics[width=0.99\textwidth]{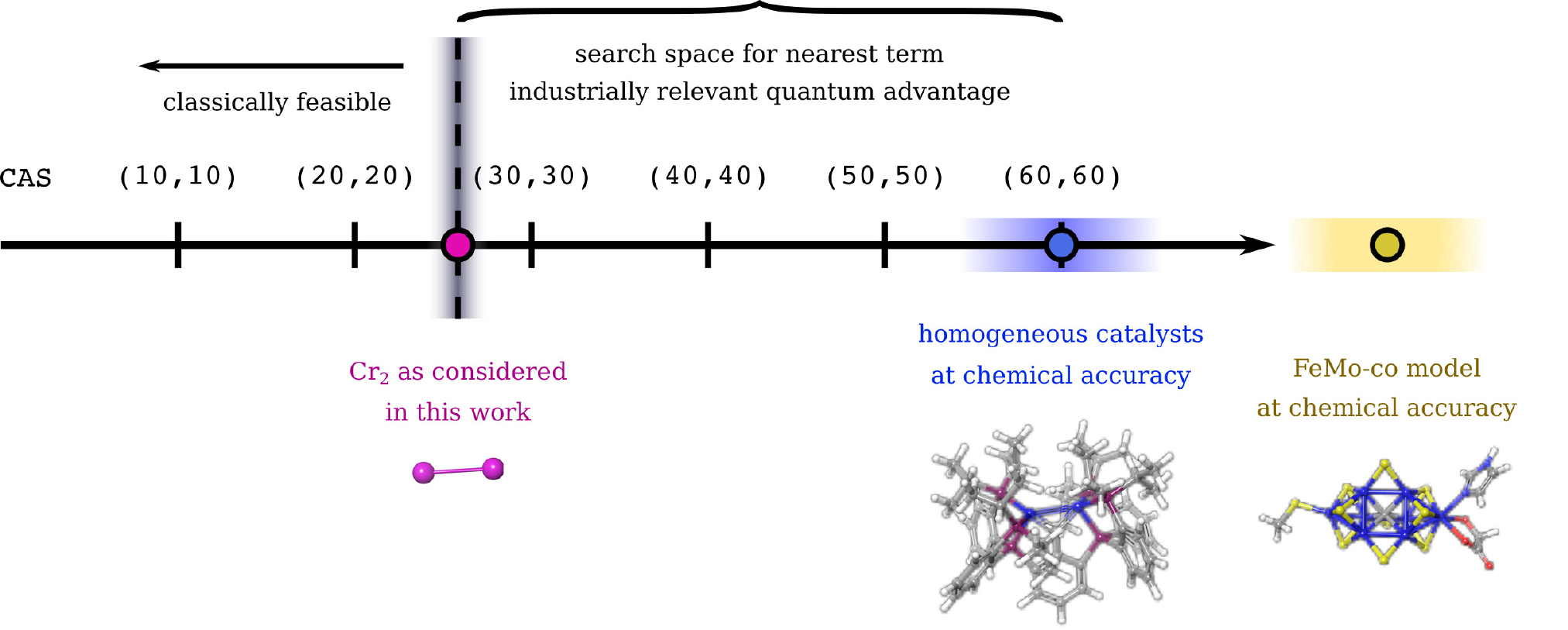}
	\begin{center}
		\begin{tabular}{|c|c|c|c|c|}
			\hline
			&  Light atom diatomics & Cr$_2$ & Homogeneous catalysts & FeMo-co model \\
			\hline
			Chemical complexity & Minimal & Medium & High & Very high \\ 
			Dynamic correlation & Medium & Medium & Difficult & Difficult \\
			Non-dynamic correlation & Easy & Medium & Difficult & Very difficult \\
			Industrial relevance & Irrelevant & Irrelevant & Highly relevant & Highly relevant \\
			Relativistic effects & Non-existent & Not likely & Not likely & Likely\\
			Protein matrix & No & No & No & Yes \\
			Molecular geometry & Accurate & Accurate & Likely accurate & Unclear \\
			Description on classical computer & Accurate & Sufficiently accurate & Not sufficiently accurate & Not sufficiently accurate  \\
			\hline
		\end{tabular}
	\end{center}
	\caption{A comparison of molecular features relevant for possible short term quantum computer applications and a schematic placement of their CASs estimated to be necessary for achieving chemical accuracy. Light atom diatomics are not put on the CAS axis because of the unclear boundaries between the dynamic and non-dynamic correlation in these molecules. 
		\label{tab:pros-cons}}
\end{table*}

We compare a single-distillation block optimized strategy and speed-optimized block strategy, both from Ref.~ \ocite{Litinski2019gameofsurfacecodes}, for both Trotterization and qubitization, in \tabref{detailed_results}. We consider error rates $p=10^{-3}$ (which has already been achieved in experiment)\cite{PhysRevLett.117.060505, Schafer2018} and $p=10^{-6}$ (`(very) long-term' hardware ambitions), and one may interpolate between the results. We find among these results several orders of magnitude in variations depending on the strategies employed. Although sparse qubitization requires about 10 times more physical qubits, the total runtime is $10^4$ times shorter than when using Trotterization. This suggests that early practice with fault-tolerant systems may use Trotterization approaches but fail to deliver early quantum advantages, while the qubitization-type approaches may yield the best results longer term. We note that the Trotterization vs qubitization discussion is ongoing and more research is required to draw concrete conclusions. For example, further research is required in order to check whether the Trotter number, and Trotter order, is sufficient or can be reduced further. It may be that these parameters are often overestimated,\cite{childs2019theory} but they certainly are hard to bound tightly for specific cases. We found that using the sparse Hamiltonian from sparse qubitization for the Trotterization protocol reduces the overall runtime by less than $30\%$, which may still be an approach to consider although the gain is not attractive compared to the sacrifice in Hamiltonian representation accuracy. One may also consider employing strategies to reduce the sub-components of the Trotter step oracle costs, like in Refs.~\ocite{wang2020resourceoptimized, pallister2020jordanwigner}. 

Here we considered only the surface code with parameter regimes typical of blueprints for future large-scale superconducting qubits based FTQC devices, which has nearest-neighbour interactions. It could be of great interest to explore other error correction paradigms and how they would impact the overhead, and in what kind of hardware these could be implemented. 

Although our resource estimates give an indication of the current state of the art in algorithms, these are not necessarily lower bounds and simulation algorithms may improve by orders of magnitude both in terms of complexity and pre-factor (overhead). With improvements in Hamiltonian simulation algorithms, reducing overhead with improved fault-tolerant quantum error correcting codes, and lower error rates in hardware with faster gate speeds, additional orders of magnitude improvements are expected over the next decade. Additionally, we stress that the real challenge is going from 10 to a million physical qubits, i.e. building a \textit{scalable} platform.\cite{Lekitsche1601540, PhysRevA.89.022317, Vinet_2018} Going from a million to a billion qubits is then a different challenge in that the scaling method has at that point been developed. All in all, it can be considered likely that these improvements will make both the scaling and the actual runtimes feasible for \textit{relevant} chemistry problems.

\section{Promising quantum chemical applications of quantum computers}
What are the types of quantum chemical applications for which quantum computers might have a clear practical advantage over classical computers in the short term? The use of quantum computers for small molecules containing only light atoms (which have been an object of study and research direction in most of the quantum computing literature so far) does not appear to be promising. Small and uncomplicated molecules like H$_2$ or LiH can in many cases be accurately treated on classical computers. Then, the processes involving only small molecules do not have a strong significance for industrial applications. Finally, the correlation energy of such molecules (the difficult part of the quantum chemistry problem) is dominated by the dynamic correlation. The only reasonable use of quantum computers for such molecules, even if the irrelevance of such calculations were ignored, is to compute their properties with at least chemical accuracy. This requires large basis sets, which in turn require a number of coherent qubits that seems impossible in the near future. The only meaningful purpose that small, light atom molecules can serve in quantum computer research is to provide a development platform, a stepping-stone toward relevant applications.

On the other side of the spectrum of possible applications of quantum computers are large and complicated molecules, the properties and chemistry of which stand little chance of being accurately described by classical computers. An example of such a molecular system is the FeMo-co protein system. Although very relevant for possible industrial applications, FeMo-co presents an overwhelming complexity for a short-term quantum computer. The smallest model of FeMo-co comprises 39 atoms, but even if that is not daunting, the large number of intermediates in the catalytic cycle to consider,\cite{Dance-IC-2011, Siegbahn-JACS-2016, Dance-I-2019} the indubitable influence of the protein matrix\cite{Benediktsson-IC-2017, Djurdjevic-CAAJ-2017, Raugei-PNAS-2018} that needs to be taken into account, the large size of minimal CAS that is expected to require 100 spatial orbitals or more, and the need to take relativistic and solvation\cite{Durrant-BJ-2001, Harris-IC-2011, Raugei-PNAS-2018} effects into consideration make it an uphill struggle for the near term. On top of this, there is no chance to describe dynamical correlation of this system on a quantum computer in the near term. So the high accuracy of description of non-dynamical correlation possibly achievable on a quantum computer is at risk of being mixed with a low accuracy description of dynamic correlation on a classical computer, making the combined result possibly untrustworthy. Finally, it is not clear whether molecular geometries of the catalytic core provided by a DFT method would be sufficiently accurate to not compromise the presumed high accuracy of the final single point energy calculations. Discussing a way to obtain energy gradients of the combined quantum-classical method seems to be premature at this point.  

We believe that for a relevant, successful application of quantum computer computations, a ``sweet spot'' lies somewhere in the middle between tiny, light atom molecules and large, staggeringly complex multi-metal active sites of protein complexes. For the middle ground we seek an application that is still relatively inaccessible to quantum computers, but not so complex as to make a \textit{tour de force}, one day achievable on a quantum computer, almost irrelevant due to the multitude of gross approximations that would have to be taken. The successful application must have relevance to real world industrial research. For this reason the chromium dimer molecule, lacking industrial applications, cannot be regarded as a good ``ultimate'' target for quantum computers. But it can be used as model or rather as a testing ground on a way to the middle ground system that we are seeking.

One type of chemistry for such a middle ground can be found among medium-sized inorganic catalyst molecules which are a subject of homogeneous catalysis research. A very recent report\cite{burg2020quantum} investigates the applicability of quantum algorithms to the quantum chemical description of a ruthenium-containing catalyst designed to convert CO$_2$ into methanol. The authors propose the treatment of active spaces that span 48-76 electrons and 52-65 orbitals. An attractive subclass of homogeneous catalysts on which early quantum advantage studies can be focused are biomimetics.\cite{Simmons-CCR-2014, Wang-CR-2017} These normally di- and tri-metal complexes borrow chemical insight from natural, metal-containing enzymes and are designed for tackling industrially important chemical transformations such as C-H bond activation or the N$_2$ bond cleavage.\cite{MacKay-CR-2004, Arnett-JACS-2020} Highly efficient C-H bond activation, for example, is at the heart of the idea of ``methanol economy'',\cite{Olah-ACIE-2005, Olah-book} which seeks to replace petroleum and coal by cleaner sources of energy and synthetic materials. 

Just like in the case of FeMo-co, currently feasible computations can guide the design and contribute to the understanding of the mechanisms of metal-containing inorganic catalysts, but they cannot replace or even eclipse a real experiment yet. However, it can be argued that quantum computers can provide a decisive advantage for modeling biomimetics and other homogeneous catalysts by helping deliver significantly more accurate computational predictions for these systems than currently possible. Importantly, these molecules are not associated with a protein matrix, which in turn tremendously reduces the complexity of the underlying chemical problem. 

In addition to containing several transition metal atoms, biomimetic catalysts usually feature a number of bulky inorganic ligands. The size and chemical complexity of these systems puts them on the verge of the capabilities of the existing quantum chemical methods. Taking Cr$_2$ for the simplest model for such dimetallic reactive centers, and considering the previous estimates for a single-metal homogeneous catalysts,\cite{burg2020quantum} we presume that the CASs sufficient for targeting chemical accuracy of small biomimetics would involve about 60 orbitals and electrons. This CAS size should be sufficient for an accurate description of their non-dynamic correlation, though their dynamical correlation would have to be dealt with separately. For a comparison of pros and cons of treating various types of chemical systems on a quantum computer see Table~\ref{tab:pros-cons}.

The homogeneous catalyst molecules considered in the previous paragraphs can seem like the smallest and simplest industrially relevant targets for early quantum advantage. However, our estimates as well as those by von Burg and co-workers\cite{burg2020quantum} indicate that their treatment will require at least thousands of logical qubits, which is a far cry from what is achievable in the nearest future. Therefore, it would be worthwhile to find still simpler targets. 

Our timing estimates for the chromium dimer (see Fig.~\ref{scaling_wallclock}) places quantum advantage for CASs of the structure (N, N) in the region after about $N \approx 26$. If we generalize this result to other types of molecules, peculiarly, and perhaps coincidentally, quantum advantage should begin after the classical quantum chemical limit which corresponds to CAS (24, 24).\cite{Jeong-JCTC-2020} So there is a gap between $N \approx 26$ after which the quantum advantage can be expected and $N \approx 60$ which is required for the accurate treatment of the non-dynamic correlation of homogeneous catalysts. Quantum advantage should play a critical role for dealing with CASs which are larger than what can be handled on a classical computer. Importantly, by virtue of their much better scaling, quantum algorithms would permit calculations on ever greater CAS sizes to make sure that the calculation will achieve a converged result. 

Are there industrially relevant chemical systems or processes, for which at least the non-dynamic correlation can be described by CAS of the type (N, N), where $26 \leq N \leq 60$? One area which may be suitable for being treated with such CAS sizes is benchmarking of efficient but approximate DFT calculations for relatively simple chemistries: calculating the error bars produced by such DFT calculations (particularly on transition states, where the stretched bonds are challenging for DFT) as well as developing corrections to account for these errors. Such a quantum-enhanced benchmarking approach may allow for more accurate prediction of reaction rates, which could have some, albeit limited, industrial relevance. Another idea for searching impactful applications that fit within the CAS range $26 \leq N \leq 60$ is a systematic processing (screening) of a comprehensive list of industrially or synthetically important chemical reactions. Selecting the minimal CAS size, sufficient for chemical accuracy, in a given structure or rate-determining step of a catalytic cycle is a difficult problem.\cite{Veryazov-IJQC-2011, Keller-JCP-2015, Stein-JCTC-2016} However, promising\cite{Aquilante-JCP-2020} automated algorithms for CAS selection have started to appear,\cite{Stein-JCTC-2016-automated, Bao-JCTC-2018, Sayfutyarova-JCTC-2019} and they could be used in an automated mining process of the industrially relevant applications, which fall within our search space (CAS with $26 \leq N \leq 60$) and are thus likely to benefit from early quantum advantage.

\section{Conclusion and outlook}

In this paper we focused on the application of quantum computational chemistry to ab initio molecular energy simulations. We found that evaluating the opportunities for noisy intermediate scale quantum devices to reach a relevant quantum advantage is not straightforward due to the difficulties associated with putting hard bounds on the performance of the variational algorithms typically employed on such devices in the presence of noise and combined with (partial) error-mitigation techniques. The first generations of fault-tolerant devices may be suitable for simulating systems with a high degree of non-dynamic correlation, as they require fewer logical qubits for quantum advantage ($10^2-10^3$) than systems with a focus on dynamic correlations ($3\cdot10^3$ and above). Although there may be ways to bring that latter number down, e.g. through the use of explicitly correlated methods. \cite{Kong-CR-2012,Sirianni-JCTC-2017, Ma-JCTC-2019,Shakhova-JPCA-2018, Townsend-JPCA-2019,Adler-JCP-2007, motta2020quantum}

Even though an exponential speedup of quantum chemical calculations is theoretically expected on quantum hardware, a significant obstacle to consider is the enormous prefactor to the polynomial runtime of quantum computational algorithms. This prefactor is partially due to the desired chemical accuracy requiring a long circuit decomposition in the gate-based model, but primarily it is due to the enormous overhead that fault-tolerant error correcting codes require. Future improved schemes, for example those exploiting better hardware connectivity or faster gates, than the superconducting qubit platform we considered here, may drastically reduce that overhead as compared to our findings.

Another important, but so-far neglected or forgotten, factor to consider is the practicality associated with routinely running quantum-chemistry simulations on quantum-processors. For instance, when discussing future applications on quantum computers, we should not discount such ``mundane'' problems as having to deal with $\sim N^4/8$ two-electron integrals. While the number of the integrals scales ``merely'' as the fourth power of molecular size, this number is nevertheless very large, amounting to $\sim 1$ terabyte of data for only 1000 orbitals. Quantum advantage might well disappear if the bandwidth for moving such amounts of data between the classical and quantum components of the quantum computer or recomputing these integrals on the fly on a classical computer becomes a computational bottleneck. This problem has recently been discussed in Ref.~\ocite{burg2020quantum}, but further research is necessary in order to understand how loading such massive data may work in practice and what can be done to improve these circumstances.

Here it appears useful to draw a parallel with the lack of widespread GPU adoption in quantum chemistry. A decade ago, GPU-enabled codes in application to quantum chemistry were regarded as a very promising\cite{Stone-JMGM-2010, Gotz-ARCC-2010} or even revolutionary\cite{Wolf-CN-2010} technology. However, despite the fact that many quantum chemical methods have been ported to GPU,\cite{walker2016electronic, Doran-JCTC-2016, Kalinowski-JCTC-2017, Tornai-JCTC-2019, Shee-JCTC-2018, Liu-IJQC-2019} with significant speedups reported, quantum chemistry production level and large scale calculations are still routinely performed on CPUs. For example, GPUs have not yet made a difference in deciphering the mechanism of FeMo-co action or in designing effective homogeneous catalysts with transition metal atoms. And GPUs were curiously absent from the account of a recent competition to produce the most accurate energy of benzene by CI-like methods.\cite{Eriksen-A-2020} 

The similarities between the advantages expected to be brought about by the algorithmic and hardware advances associated with GPUs and quantum computers tell us to pay close attention to the lessons learned from the lack of widespread GPU adoption in quantum-chemistry. For instance, efficient implementation of quantum chemistry algorithms on GPU required a total redesign of the conventional quantum chemistry codes.\cite{Titov-JCTC-2013, Tornai-JCTC-2019} Therefore efficient GPU codes were not as widely available as efficient CPU codes. The speedups for GPU calculations using basis sets with larger angular momenta and in other types of calculation were quite modest.\cite{Yasuda-IJQC-2014, Morrison-JCC-2018, Mullinax-JCTC-2019} Also, competition for GPU resources from other application fields, like molecular mechanics methods (a field in which GPU use proved to be truly transformative) was likely a strong factor. And even the higher dollar cost of specialized GPU hardware in comparison with more versatile CPU hardware played a role.

We surveyed the types of molecular systems and quantum chemical applications that are likely to display a \textit{relevant} quantum advantage, when quantum computer hardware comes of age. Our discussion was partly stimulated by the intention to review and reformulate the popular belief that quantum chemistry calculations constitute a ``low-hanging fruit'' for quantum computers. The molecular systems often mentioned in the quantum computing literature range from diatomics to biological molecules.\cite{Outeiral-WIRES-2020} Because of the indiscriminate nature of these chemical systems thus discussed, it is perhaps assumed that any kind of molecular system could be of interest to quantum computer calculations, as there must be a limit to how fast and how accurately any of them can be described on a classical computer. However, our analysis shows that only certain types of quantum chemistry problems are projected to benefit from quantum advantage. For targets of near term quantum computations we propose molecular systems of intermediate complexity. Ideally, they would have a significant non-dynamic component in their correlation energy, and would be treatable accurately with CAS (N,N) where $26 \leq N \leq 60$. These targets should offer a sufficient challenge and an ample vista of real world systems that should be aimed at in a more distant future.

\bibliographystyle{achemso}
\bibliography{references}

\end{document}